\documentclass[aps,prl,floatfix,twocolumn,showpacs]{revtex4}
\usepackage{graphicx} \usepackage{amsmath} \usepackage{amssymb}
\usepackage{color}
\newcommand{\comment}[1]{}
\newcommand{\BEQ}{\begin{equation}}
\newcommand{\EEQ}{\end{equation}}
\newcommand{\BEA}{\begin{eqnarray}}
\newcommand{\EEA}{\end{eqnarray}}

\renewcommand{\d}{{\rm d}}

\newcommand{\ssum}{{\sum}}
\newcommand{\pprod}{{\prod}}

\newcommand{\half}{\frac{1}{2}}

\renewcommand{\a}{\alpha}

\newcommand{\g}{\gamma}
\newcommand{\bJ}{\bar{J}}
\newcommand{\bs}{\bar{s}}
\newcommand{\baf}{\bar{f}}
\newcommand{\at}{{\rm atanh}}
\begin{document}
\bibliographystyle{unsrt}

\title{Community Detection with and without Prior Information}

\author{ Armen E. Allahverdyan$^1$, Greg Ver Steeg$^2$,  and Aram Galstyan$^2$}
\affiliation{$^1$Yerevan Physics Institute,
Alikhanian Brothers Street 2, Yerevan 375036, Armenia,\\
$^2$Information Sciences Institute, University of Southern California, 
Marina del Rey, CA 90292, USA}

\begin{abstract} We study the problem of graph partitioning, or
clustering, in sparse networks with prior information about the
clusters. Specifically, we assume that for a fraction $\rho$ of the nodes
 their true cluster assignments are known in advance. This can be
understood as a semi--supervised version of clustering, in contrast to
unsupervised clustering where the only available information is the
graph structure. In the unsupervised case, it is known that there is a
threshold of the inter--cluster connectivity beyond which clusters
cannot be detected.  Here we study the impact of the prior information
on the detection threshold, and show that even minute [but
generic] values of $\rho>0$ shift the threshold downwards to its lowest possible
value. For weighted graphs we show that a small semi--supervising can
be used for a non-trivial definition of communities.
\end{abstract}


\maketitle

Graph partitioning is an important problem with a wide range of
applications in circuit design, data mining, social sciences, {\it etc}
\cite{newman}.  In the context of social network analysis, a relaxed
version of this problem is known as {\em community detection}, where
community is loosely defined as a group of nodes so that the link
density within the group is higher than across different groups.  Many
real--world networks have well--manifested community
structure~\cite{newman}, which explain the significant attention this
problem has received recently.  Indeed, much recent research has focused
on developing community detection methods using various approaches. A
recent review of existing approaches can be found in \cite{fortunato}. 

Generally, most algorithms are able to detect communities accurately if
the number of inter--community edges is not very large. The
detection becomes less accurate as one increases the density of links
across the communities. In fact, most community detection algorithms
seem to have an intrinsic threshold in inter--community coupling beyond
which detection accuracy is very poor~\cite{danon}.  Recently this
problem has been studied theoretically by formulating community
detection as a minimization of a certain Potts-Ising Hamiltonian
\cite{leone}. It was shown that the graph partitioning problem is indeed
characterized by a phase transition from detectable to undetectable
regimes as one increases the coupling strength between the clusters
\cite{leone}. Specifically, for sufficiently large inter--cluster
coupling, the ground state configuration of the Hamiltonian has random
overlap with the underlying community structure.

Most work on community detection so far has considered {\em
unsupervised} version of clustering, where the only available
information is the graph structure.  In many situations, however, one
might have additional information about possible cluster assignments of
certain nodes. Generally speaking, such information can be in form of
pair-wise constraints (via must-- and cannot links), or, alternatively,
via known cluster assignments for a fraction of nodes.  Here we consider
the latter scenario, which has attracted recent interest in the context
of semi--supervised learning and classification~\cite{zhu,getz,leone_sumedha}. For instance, classification of text documents can be posed as a graph clustering problem, with links based on proximity for some similarity score. In this case, we could ask how picking some small random fraction of documents to be classified by humans will affect our clustering algorithm. Semi--supervised learning falls in between unsupervised
(i.e., regression) and totally supervised methods (i,e., clustering).
The main premise of the semi--supervised learning is to use prior
information about fraction of data points in order to facilitate the
classification of the other nodes. Here we are specifically interested
in graph--based semi--supervised learning. In this approach, one first
maps the data to a (weighted) graph using pair--wise similarities between
different data points, and then partitions the nodes in this the graph,
e.g., using spectral clustering. We note that while most clustering methods have been
developed for unweighted (homogeneous) networks, generalization 
to the weighted situation has been suggested as well~\cite{fortunato,newman_weighted}.

Despite extensive amount of work in semi--supervised learning in recent
years, there is a lack of adequate theoretical development. The purpose
of this Letter is to present a theoretical analysis of the
semi--supervised version of the community detection, and uncover new
scenarios of community detection facilitated by semi--supervising.  Here
we focus on the so called {\em planted bisection} graph
model~\cite{karp,leone}, where the clusters are introduced by hand
(implanted), and one checks whether the clustering method under
consideration will recognize them.  This model is the (supposedly)
simplest laboratory for studying the foundations of clustering methods. 

Our main contributions can be summarized as follows: For unweighted
graphs, we show analytically that any small (but finite) amount of prior
information destroys the critical nature of cluster detectability, by
shifting the detection threshold to its lowest possible value.
Furthermore, for graphs where links within and across communities have
different weights, we find that the semi--supervision leads to
detectable clusters even below the intuitive weigh--balanced value. Note
that for weighted graphs the very definition of the communities is rather
ambiguous. Our results suggests that the availability of prior
information might resolve this ambiguity.

{\em Model}: Consider an Erd\"{o}s--R\'{e}nyi graph where each pair of nodes
is linked with probability $\a/N$, and where $N$ is the number of nodes in
the graph. We assume that each link carries a weight $J>0$. Now imagine a pair of such
identical Erd\"{o}s--R\'{e}nyi graph, which models two clusters
(communities). Besides the intra-cluster $J$-links, each node in one
graph is linked with probability $\g/N$ with any node of another
Erd\"{o}s--R\'{e}nyi graph.  These inter-cluster nodes are given weight
$K>0$. For clarity, both $J$ and $K$ are assumed to be integer numbers. 

This {\em planted bisection} graph model~\cite{karp} will be employed for studying the
performance of the cluster detection method, which places an
Ising spin on each node and lets these spins interact via the network
links~\cite{reich,leone}:
\BEA
\label{1}
H =- \ssum_{i<j}^N J_{ij}s_is_j - \ssum_{i<j}^N\bJ_{ij}\bs_i\bs_j- \ssum_{i,j}^NK_{ij}s_i\bs_j,
\EEA
where we made the bi--cluster nature of the network explicit by
introducing separate spin variables $s_i=\pm 1$ and $\bs_i=\pm 1$
($i=1,\ldots,N$) for two clusters. Here
$J_{ij}$ and $\bJ_{ij}$ are identically and independently distributed
random variables which assume zero with probability $1-\frac{\a}{N}$ and
$J>0$ with probability $\frac{\a}{N}$. Likewise, $K_{ij}$ identically
and independently are equal to zero with probability $1-\frac{\g}{N}$
and to $K>0$ with probability $\frac{\g}{N}$. 
In order to enforce equipartition, the Hamiltonian (\ref{1}) will be studied under the constraint
\BEA
\label{2}
\ssum_{i=1}^Ns_i+\ssum_{i=1}^N\bs_i =0.
\EEA
Thus, detecting the sign of a given spin $s_i$ at zero temperature
(so as to exclude all thermal fluctuations) we can conclude to
which cluster the corresponding node belongs: all spins having equal signs belong to the same cluster. The error probability for the cluster assignment is $p_{\rm e}$, which can also be viewed as the fraction of incorrectly identified spins or one minus the probability of correctly identifying a node's community. This error is directly related to magnetization,
\BEA
p_{\rm e}&=& (1-|m|)/2, \\
 m & \equiv & [\langle s_i\rangle_{T=0}]_{\rm av}=-[\langle \bs_i\rangle_{T=0}]_{\rm av},
\label{prisma}
\EEA
where $m$ is the (single--cluster) magnetization, $\langle \ldots \rangle_{T=0}$ is the
zero-temperature Gibbsian average, i.e.  the average over all
configurations of spins having in the thermodynamic limit the minimal
energy given by (\ref{1}, \ref{2}), and where $[\ldots ]_{\rm av}$ is the average over the bi-graph
structure, i.e., over $\{J_{ij}\}$, $\{\bJ_{ij}\}$ and $\{K_{ij}\}$. As
implied by the self-averaging feature, instead of taking $[\ldots ]_{\rm
av}$, we can evaluate $\langle s_i\rangle_{T=0}$ on the most probable
bi-graph structure(s).

The above formulation refers to the unsupervised community detection. The semi--supervising implies  that for some  nodes
their cluster assignment is known in advance~\cite{getz,leone_sumedha}. Here we assume that these nodes are distributed randomly over the graph. To account for the prior information, we introduce infinitely strong magnetic fields acting on those node in the appropriate direction. Thus, the Hamiltonian (\ref{1}) is
modified as follows: 
\BEA
\widetilde{H}=H
-\ssum_{i=1}^N f_is_i  -\ssum_{i=1}^N \baf_i\bs_i ,
\label{1.1}
\EEA
where 
$f_i$ (resp. $\baf_i$) are identically and independently distributed
random variables that are equal to $0$ with probability $1-\rho$ and to
$\infty$ (resp. $-\infty$) with probability $\rho$. The
constraint (\ref{2}) is satisfied in the average sense. Thus, with respect of
two randomly chosen sets of spins (each containing $\rho N$ members) we
know exactly to which cluster they belong, since $f_i=\infty$ implies $s_i=1$. Below we study the 
threshold of cluster detection with and without semi--supervising.

It is known that Ising models on Erd\"{o}s--R\'{e}nyi graphs can be
efficiently studied via the cavity method; see, e.g,
\cite{mezard,kanter,gold_rsb}. The main object of this method is the
probability $P(h)$ of an internal field acting on one $s$-spin. The
physical order-parameters are expressed as moments of $P(h)$; see
(\ref{bela}, \ref{bartok}). As applied to our Hamiltonian (\ref{1.1}),
the cavity method produces the following equation for $P(h)$ [the
derivation of this equation is fairly similar to that given in
\cite{mezard,kanter,gold_rsb} for the ordinary Ising model on the
Erd\"{o}s--R\'{e}nyi graph; so we shall repeat it here]:
\BEA
P(h)= \ssum_{n=0}^\infty \ssum_{m=0}^{\infty}  \frac{\a^n e^{-\a}}{n!}\,\frac{\g^m e^{-\g}}{m!}
\times~~~~~~~~~~~~~~~~\nonumber\\ 
\int\hat{p}(f)\d f
\int\pprod_{k=1}^m P(h_k)\d h_k\,\,\int\pprod_{l=1}^n \bar{P}(g_l)\d g_l\times \nonumber\\
\delta  \left(\, h -f 
- \ssum_{k=1}^m \phi[h_k,J]
 - \ssum_{k=1}^n \phi[g_k,K]  \,\right),
\label{hatkhi}
\EEA
Here we already assumed the zero-temperature limit \cite{dvin} denoted
$\phi[a,b]\equiv{\rm sign}(a)\,{\rm min}[\, |a|, b\,] $ and 
$g_k$ (resp. $h_k$) are the fields acting on the $s$-spin from
$\bs$-spin (resp. from other $s$-spins). These fields naturally enter
with weight $\frac{\g^m e^{-\g}}{m!}$ (resp. $\frac{\a^n e^{-\a}}{n!}$),
which is the {\em excess} degree distribution of the corresponding
Erd\"{o}s--R\'{e}nyi network. 

In (\ref{hatkhi}), the distribution of the frozen
(supervising) field acting on $s$-spins is determined by
\begin{equation}
 \hat{p}(f)=\rho\delta(f-\infty)+(1-\rho)\delta(f).
\end{equation}
  Due to (\ref{1}--\ref{1.1})
and the complete inversion symmetry between the two clusters, we can
take $\bar{P}(g)=P(-g)$, and then (\ref{hatkhi}) is worked out via the
Fourier representation of the delta-function yielding 
\begin{equation}
P(h)=\rho
\delta(h-\infty)+(1-\rho)\widetilde{P}(h),
\label{fantom}
\end{equation} 
where $\widetilde{P}(h)$ refers to those $s$-spins, which were not directly frozen by infinitely
strong random fields. 

It can be seen from (\ref{hatkhi}, \ref{fantom}) that $\widetilde{P}(h)$ satisfies the following equation:
\BEA
\widetilde{P}(h)=e^{-\a-\g}\int\frac{\d z}{2\pi}\, e^{iz h}\, \exp \biggl [    \a \rho\, e^{-izJ}+\g \rho\, e^{izK}  \nonumber \\
 +\a(1-\rho) \int \d g_1 \widetilde{P}(g_1) e^{-iz\,{\rm sign}(g_1)\,{\rm min}[|g_1|, J]} \nonumber \\
 +\g (1-\rho)\int \d g_2 \widetilde{P}(g_2) e^{iz\, {\rm sign}(g_2)\,{\rm min}[|g_2|, K]} \biggr ].
\label{tartar}
\EEA
The physical order-parameters are expressed as
\cite{mezard}: 
\BEA
\label{bela}
m=[\,\langle s_i\rangle_{T=0}\,]_{\rm av} =\int\d h\, \widetilde{P}(h) \, {\rm sign}(h),~~~~~~~~~~\\
q=[\,\langle s_i\rangle_{T=0}^2\,]_{\rm av}
=\int\d h\, \widetilde{P}(h) \, {\rm sign}^2(h), ~~ {\rm sign}(0)=0,~~
\label{bartok}
\EEA
where $[\ldots]_{\rm av}$ is now the average over the bi-graph structure and the random fields.
Recall that $m$ defines the error probability according to (\ref{prisma}).
In (\ref{bartok}) $q$ differs from $1$ due to possible
contribution $\propto \delta(h)$ in $\widetilde{P}(h)$. Thus, $1-q$ is the
fraction of spins that do not have definite magnetization, since they do
not belong to the sub-graph of strongly connected spins [which exists
above the percolation threshold], while $q-m$ is the fraction of spins
that do not have definite magnetization, because they are strongly
frustrated, though they do belong to the sub-graph of strongly connected
spins. 
 
 {\it Unweighted} ($J=K$) {\it unsupervised} ($\rho=0$) situation:
Since at $T=0$ only the ratio $J/K$ matters (and not the absolute values
of $J$ and $K$) we assume $J=K=1$. Now the local fields can attain only
integer values and the solution of (\ref{tartar}) is searched for as
\BEA
\label{artashir}
\widetilde{P}(h)=\ssum_{n=-\infty}^\infty c_n \delta(h-n),
\EEA
which upon substituting into (\ref{tartar}) and using
\BEA
\label{boost}
e^{\frac{x}{2}(e^{y+iz}+e^{-y-iz})}  = \ssum_{n=-\infty}^\infty I_n(x)e^{-ny-inz},
\EEA
where $I_n(x)$ is the modified Bessel function, produces
\begin{gather}
\label{baran}
c_n = e^{-(\alpha+\gamma)q-ny}I_n(x),
\end{gather}
where
\begin{gather}
\label{baran2}
x\equiv\sqrt{(\a+\g)^2q^2-(\a-\g)^2m^2}, \nonumber \\ 
y\equiv \at\frac{(\g-\a)m}{(\a+\g)q}.\nonumber
\end{gather}
This then implies via (\ref{bela}, \ref{bartok}, \ref{artashir})
\begin{align}
\label{ha}
&1-q = e^{-(\alpha+\gamma)q} I_0[x], \\
& m=-2e^{-(\alpha+\gamma)q}\ssum_{n=1}^\infty I_n(x)\sinh [ny].
\label{che}
\end{align}
Eq.~(\ref{che}) predicts a second-order transition, where $m$ is the order-parameter.
In the vicinity of the second-order phase transition one can expand (\ref{ha}, \ref{che})
over $m$:
\BEA
\label{gr}
1-q=e^{-(\alpha+\gamma)q} I_0[(\alpha+\gamma)q], \\
1=(\alpha-\gamma)(1-q)\left(
1+\frac{I_1[(\alpha+\gamma)q]}{I_0[(\alpha+\gamma)q]}
\right).
\label{at}
\EEA
where we employed identities involving Bessel functions. 
We have $m>0$ ($m=0$) if the RHS of (\ref{at}) is larger (smaller) than its LHS.

\begin{figure}[ht]
\vspace{0.6cm}
\center
\includegraphics[width=7.5cm]{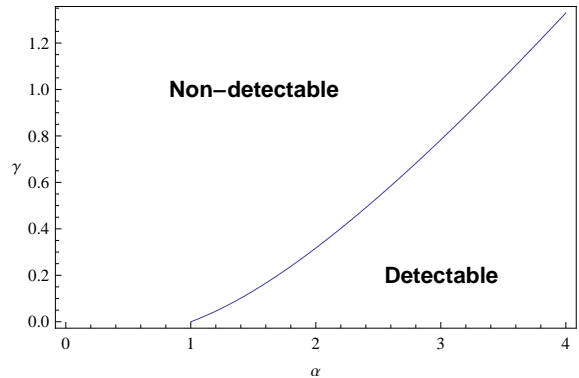}
\vspace{0.4cm}
\caption{ The phase diagram for $J=K=1$. The line on the $(\a,\g)$ plane indicates
second-order phase-transition from $m=0$ (no clustering detection) to $m>0$ (clustering detection). } 
\label{f1}
\end{figure}

Eq.~(\ref{at}) determines the detection threshold, above of which
the method is capable of detecting clustering with {\em better than random} probability of error~(\ref{prisma}). In the $\a-\g$ plane, the threshold line starts from $(\alpha=1,\gamma=0)$, see Fig.~\ref{f1},
since (\ref{gr}) predicts a percolation bound for $q$: $q=0$ ($q>0$) for
$\a+\g<1$ ($\a+\g>1$). Naturally, close the percolation bound
$\alpha=1$, even very small inter--cluster coupling $\g$ nullifies $m$.
Fig.~\ref{f1} shows that at the detection threshold $\a>\g$;
moreover the difference $\a-\g$ at the threshold grows as
$\sqrt{2\pi(\a+\g)}$ for a large $\a+\g$; see (\ref{gr}).  Thus, the
ratio $\frac{\a-\g}{\a+\g}$ converges to zero for a large $\a+\g$. In
this {\it weak} sense, the detection threshold converges to $\a=\g$ for
 large $\a+\g$, while for any finite $\a$ the unsupervised clustering
detection threshold lies below the line $\a=\g$; see Fig.~\ref{f1}. 

\begin{figure}[ht]
\vspace{0.6cm}
\center
\includegraphics[width=7.5cm]{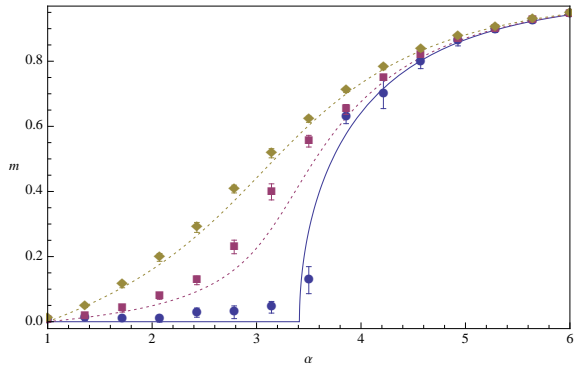}
\vspace{0.4cm}
\caption{ Normal curve: magnetization $m$ versus $\alpha$ for $\gamma=1$. $m$ undergoes second-order phase-transition at $\a=3.4$.
Dashed curves: remnant [semi-supervised] magnetization 
$m$ versus $\alpha$ for $\gamma=1$ for  $\rho=0.2$ (top) and $\rho=0.05$ (bottom). Symbols: mean magnetization (with $99\%$ confidence interval) for numerical experiments using simulated annealing on graphs with 10,000 nodes for $\rho=0$ (circles), $0.05$ (squares), and $0.2$ (diamonds). }
\label{f2}
\end{figure}
 
{\it Under semi-supervising} we still employ (\ref{artashir}, \ref{boost}) 
and obtain (\ref{baran}, \ref{ha}, \ref{che}), but now in the
RHS of these equations one should substitute
$m\to\rho+(1-\rho)m$ and $q\to\rho+(1-\rho)q$.
Expanding over a small $m$ we get
\BEA
\label{haha}
1-q=e^{-(\alpha+\gamma)(\rho+[1-\rho]q)} I_0[(\alpha+\gamma)(\rho+[1-\rho]q)],~~ \\
m=\rho(\alpha-\gamma)(1-q)\left[
1+\frac{I_1[(\alpha+\gamma)(\rho+[1-\rho]q)]}{I_0[(\alpha+\gamma)(\rho+[1-\rho]q)]}
\right]. \nonumber
\EEA
Now $m>0$ for any $\alpha-\gamma>0$. This is the average-connectivity
threshold, which for the considered unweighted scenario is the only
possible definition of clustering. Thus, {\it any} generic semi-supervising 
leads to the theoretically best possible threshold $\a=\g$; see Fig.~\ref{f2}. 

Note that for $\rho>0$, (\ref{haha})
has a non-trivial solution $q<1$ for arbitrary
$\alpha+\gamma>0$: the percolation bound is also diminished by a
small [but generic] $\rho$. 

\begin{table}
\begin{tabular}{|c||c|c|c|c|c|c|} 
\hline
~$\rho$~ &    $0.3$  &  $0.1$  & $0.005$  & $0$  \\
\hline
\hline
~$\a$~   & $1.6812$  & $1.5976$   & $1.5173$  & $4.9122$    \\
\hline
~$q$~   & $0.86734$ & $0.7871$   & $0.7087$  & $0.8373$   \\
\hline
~$m_1$~  & $0.0241$  & $0.0109$   & $7\times 10^{-4}$  & $0$    \\
\hline\hline
$m|_{\alpha=2}$  & $0.0401$ & $0.0182$  & $0.0016$  & $-$\\
\hline 
\end{tabular}
\caption{Weighted situation: $2J=K=2$. For $\gamma=1$ and various semi-supervising degrees $\rho$
we list the clustering threshold $\alpha$ and the values of $q$ and $m_1$ at this threshold. 
}
\label{tab1}
\end{table}


{\it Arbitrary degree distributions} can also be considered in this framework. First, note that the excess degree distribution for $n$ intra--cluster and $m$ inter--cluster edges in (\ref{hatkhi}) can also be written in terms of an overall excess degree distribution and a parameter $p_{out}$ which determines the probability that a given edge connects to a node outside the cluster of the current node.
\BEA
  \frac{\a^n e^{-\a}}{n!}\,\frac{\g^m e^{-\g}}{m!}
= \sum_{s=0}^\infty  q(s) 
 \sum_{k=0}^s  {s \choose k} p_{out}^k (1-p_{out})^{s-k} \\
  \times \delta_{n,s-k} \delta_{m,k}. \nonumber
\EEA
In this case, $q(s) = e^{-(\g+\a)} (\g+\a)^s / s!$ and $p_{out} = \g/(\a+\g)$. Now, we want to consider $q(s)$ to be an arbitrary excess degree distribution, while $p_{out}$ measures the connection between clusters; $p_{out}= 0$ indicates disjoint clusters while $p_{out}= \half$ indicates no cluster structure. We employ a second trick by rewriting $q(s)$ in terms of a generating function $G(x) = \sum_s x^s q(s)$ using the inverse formula $q(s) = \lim_{x \rightarrow 0} \partial_x^s G(x)/s!$.

This transformation leads to expressions similar to (\ref{baran}). For $\rho=0$,
\BEA
c_n =  \lim_{x \rightarrow 0}  \left( \frac{q+\widetilde{m}}{q-\widetilde{m}}\right)^{\frac{n}{2}} I_n[  \sqrt{q^2-\widetilde{m}^2} ~ \partial_x]  ~e^{(1-q) \partial_x} ~G(x),
\EEA
where $\widetilde{m} = (1 - 2 p_{out})m$.
If we use the generating function for the Poisson distribution $G(x) = e^{-(\a+\g)(1-x)}$ and consider a power series representation of this expression, we see that we should just replace $\partial_x \rightarrow (-\a-\g)$ and we recover (\ref{baran}) exactly. 

Rewriting the modified Bessel function in integral form and using the identity $ \lim_{x\rightarrow0} e^{y \partial_x} G(x) = G(y)$, provides the following succinct expression.
\BEA
1-q &=& \frac1\pi \int_0^\pi d\theta ~G(\sqrt{q^2-\widetilde{m}^2} ~\cos(\theta) + 1-q) \\
m &=& \frac{\widetilde{m}}{q~\pi} \int_0^\pi d\theta ~\frac{G(\sqrt{q^2-\widetilde{m}^2} ~\cos(\theta) + 1-q)}{\sqrt{1-\left( \frac{\widetilde{m}}{q} \right)^2}~\cos{\theta} - 1} 
\EEA
However, depending on the integral, it may be more tractable to work with a power series representation of the Bessel functions. 
For a specific generating function, these integrals play the role of (\ref{ha}, \ref{che}), implicitly specifying $m,q$. We also recover the result that these equations apply in the supervised case with the substitutions $m\to\rho+(1-\rho)m$ and $q\to\rho+(1-\rho)q$ on the RHS. For a power--law degree distribution, we get results qualitatively the same as depicted in Fig.~\ref{f2}, replacing the $\a$ with $p_{out}$ on the $x$ axis. 
Specifically, for power--law networks, both the analytic results above and numerical experiments using simulated annealing confirm the existence of a detection threshold in the unsupervised case, while the supervised case leads to nonzero magnetization whenever $p_{out} < \half$.


{\it The weighted situation} $J\not =K$ will be studied via two
particular (but important) cases $2J=K=2$ and $2K=J=2$, to make them amenable to
analytic approach. Putting it into (\ref{hatkhi}) and using (\ref{boost}) two times, we
see that there are now four order-parameters:
\BEA
m, \qquad q,\qquad 
q_1\equiv c_1+c_{-1}, \qquad m_1\equiv c_1-c_{-1}.
\label{dum2}
\EEA
Note that only
$q$ and $m$ are observed from the single-spin statistics, see (\ref{bela}, \ref{bartok});
$q_1$ and $m_1$ can be observed only via measuring the 
internal field distribution $\widetilde{P}(h)$.

For $2J=K=2$ we introduce the following notations
\begin{gather}
C_p(q,m)= a\ssum_{n=-\infty}^{\infty}I_n(\tilde{u})I_{p-2n}(\tilde{x})
\cosh[2\tilde{y}n-\tilde{v}n-\tilde{y}p]\nonumber,\\
S_p(q,m)= a\ssum_{n=-\infty}^{\infty}I_n(\tilde{u})I_{p-2n}
(\tilde{x})\sinh[2\tilde{y}n-\tilde{v}n-\tilde{y}p],\nonumber\\
a = 2e^{-(\a+\g)q},~ z^{\pm}=q\pm m, ~ z^{\pm}_1=q_1\pm m_1,~
\xi=\frac{2\rho}{1-\rho},
\nonumber\\
\label{andes0}
\tilde{x}= (1-\rho)\sqrt{(\a z^-+\g z^+_1)(\a [z^+ +\xi] + \g z_1^- )},\\
\label{andes1}
\tilde{y}=\frac{1}{2}\ln\frac{\a z^- +\g z_1^+}{\a [z^+ +\xi] +\g z_1^-  },
~~
\tilde{v}=\frac{1}{2}\ln\frac{z^+ - z_1^+ + \xi }{z^- - z_1^-}, \\
\tilde{u}=\g(1-\rho)\sqrt{(z^- - z^-_1)(z^+ - z_1^+ +\xi)},
\label{andes2}
\end{gather}
and write down the order-parameter equations:
\begin{gather}
\label{bar}
q=\ssum_{p=1}^\infty C_p(q,m), \qquad q_1=C_1(q,m),\\
m=\ssum_{p=1}^\infty S_p(q,m), \qquad m_1=S_1(q,m).
\label{kokhba}
\end{gather}
Eqs.~(\ref{bar}, \ref{kokhba}) apply also for $2K=J=2$, but now in
(\ref{andes0}--\ref{andes2}) we should interchange $\a$ and $\g$, and
then substitute $\tilde{y}\to -\tilde{y}$ and $\tilde{v}\to -\tilde{v}$.

{\it Discussion.}
Eqs.~(\ref{bar}, \ref{kokhba}) predict a second-order transition over
$m$ and $m_1$. Similarly to the unsupervised case, the threshold of this transition 
is found via expanding (\ref{bar}, \ref{kokhba}) over $m$ and $m_1$. But
the real qualitative differences between weighted and unweighted
situations show up under semi-supervising, which we consider in more details below. 

First we focus on $2J=K=2$ and recall that the clustering threshold is
defined via $m=0$. While for the previous unweighted situation, any
amount of semi-supervision (as quantified by $\rho$) sufficed for
shifting the clustering threshold to a $\rho$-independent value, here
the detection threshold starts to depend on $\rho$, and the smallest
threshold is achieved for $\rho\to 0$; see Table 1 for an illustration. To understand this seemingly counterintuitive observation, note that detection
threshold is achieved as a balance between the inter--cluster links with the average
connectivity $\g$ and weight $K=2$, and intra--cluster links
with the average connectivity $\a$ and weight $J=1$. Consider now the impact of the semi--supervision on a test-spin. The inter--cluster links will exert negative fields on this test--spin, while the intra--cluster links will exert positive fields. Since inter-cluster links have twice larger weigh,  increasing $\rho$ facilitates the negative fields. This explains why vanishing semi-supervising $\rho\to
0$ results in a lower detection threshold. 

Another interesting observation is is that|contrary to the unsupervised situation, where $m$
and $m_1$ simultaneously turn to zero at the threshold|we get $m_1>0$ at
the semi-supervised threshold. Thus, some memory about the clustering is
conserved despite the fact that $m=0$; see Table 1.  Since $m_1$
cannot be observed via a single spin, this memory is hidden. The reason
of $m_1>0$ is that $m_1$ counts the internal fields equal to $\pm 1$,
and there are more such fields coming from the intra--cluster
[connectivity $\a$, weight $1$] links that exert positive fields due to
the semi-supervised (frozen) spins. 

Now consider perhaps the most paradoxical aspect of the semi-supervised detection
threshold: it is {\it smaller} than the value deduced from balancing the cumulative weights of
intra--cluster and inter--cluster  links, which yields  $\a J = \g K$. Indeed, according to Table
\ref{tab1} (where $\g=1$) we have $\a=1.5$ (reached for $\rho\to 0$)
versus the weight-balancing value $\a=2$. This result seemingly
contradicts the intuition we got so far: {\it i)} a rough intuition
about Hamiltonian (\ref{1}) is that it is based on defining a cluster
via the intra--cluster weight being larger than the inter--cluster weight.
{\it ii)} The unsupervised threshold is well above the weight-balancing
prediction; see Table \ref{tab1}. {\it iii)} In the unweighted case
($J=K$) the semi-supervising just reduces the detection threshold
towards $\a=\g$, which coincides with the weight-balancing value. 

To understand this effect, we turn to the physical picture of the
threshold, where positively and negatively acting links driven by the
semi-supervised (frozen) spins compensate each other. At the
weight-balance $\a J=\g K$ (with $J<K$) fewer (but stronger)
inter--cluster links have the same weight as more numerous (but weaker)
intra--cluster links. Since the intra--cluster links  are more numerous, their
overall effect on a (randomly chosen) test spin is more deterministic and hence capable
of building up a positive $m$ at $\a J=\g K$. Thus, the actual threshold
is reached for $\a J<\g K$. 

\begin{table}
\begin{tabular}{|c||c|c|c|c|c|c|}
\hline
~$\rho$~ &    $0.005$  & $0.1$ & $0.4$  & $0$  \\
\hline
\hline
~$\a$~  & $0.6706$ & $0.6481$  & $0.6259$  & $2.3922$    \\
\hline
~$q$~  & $0.5099$ & $0.6197$  & $0.7755$  & $0.8163$   \\
\hline
~$m_1$~  & $-0.002$ & $-0.0284$  & $-0.0709$  & $0$    \\
\hline
\end{tabular}
\caption{The same as for Table \ref{tab1}, but for the weighted situation: $J=2K=2$. 
}
\label{tab3}
\end{table}


We thus conclude that for weighted graph $K>J$ a small [but
generic] semi-supervising can be employed for defining the very
clustering structure. This definition is non-trivial, since it performs
better than the weight-balancing definition. Indeed, for a weighted
network the definition of detection threshold is not clear {\it a
priori}, in contrast to unweighted networks, where the {\it only}
possible definition goes via the connectivity balance $\a=\g$. To
illustrate this unclarity, consider a node connected to one cluster via
few heavy links, and to another cluster via many light links. To which
cluster this node should belong {\it in principle}? Our answer is that
the proper cluster assignment in this case can be defined via semi-supervising. 

It is interesting to calculate $m$ at the weight-balancing
value $\a J=\g K$, since this is the semi-supervising benefit of those
who would insist on the weight-balancing definition of the
threshold; see Table \ref{tab1}. Note finally that for large values of $\gamma$ both unsupervised and
semi-supervised thresholds converge to $\a J=\g K$, since
now fluctuations are irrelevant from the outset. 

All these effects turn upside-down for $2K=J=2$; see Table \ref{tab3}.
Now the threshold is minimized for the maximal semi-supervising $\rho\to
1$, $m_1$ is negative at the threshold|and thus the memory about the
clustering is contained in $m-m_1>0$|and the semi-supervised detection
threshold $\alpha$ is always larger than the weight-balancing value
$\g K/J$. These results are explained by "inverting" the above
arguments developed for $J<K$. 

{\it In conclusion}, we analyzed the community detection in
semi--supervised settings, where one has prior information about the
community assignments of certain nodes. We showed that for the planted
bisection graph model with intra--cluster and inter--cluster average
connectivities $\a$ and $\g$, respectively, even a tiny (but finite)
semi-supervising shifts the detection threshold to
its intuitive value $\a = \g$.  We observed a similar effect of lowered
detection threshold for weighted graphs. In contrast to the unweighted
case, the shift in this case depends on the degree of supervision.
Furthermore, we found that when approaching the unsupervised limit by
having $\rho \rightarrow 0_{+}$, the detection threshold converges to a
value lower (better) from the one obtained via balancing intra--cluster
and inter--cluster weights. We suggest that this can serve as an
alternative definition of clusters. We also saw that in the
semi-supervised case some (hidden) memory on the clustering survives 
at the detection threshold.  

Although this work focused on the analytically simpler case of Erd\"{o}s--R\'{e}nyi graphs, we have repeated the analytic and numerical analysis for power--law graphs  and found similar results, suggesting that the impact of the network topology is 
quantitative rather than qualitative. A similar picture has been observed in~\cite{leone}. 


An interesting generalization is to consider choosing frozen spins  more deliberately (e.g., based on node connectivity) as  opposed to random selection studied here. While this difference might not be important for ER graphs, it might lead to some significant quantitative changes for power--law graphs with a large number of well--connected hubs. Finally, it would be interesting to 
to consider prior information not only about nodes, but also about links in the network. An example is a constraint that two nodes 
belong to the same community. In our model, this can be incorporated by making the coupling between those pairs sufficiently strong. This scenario resonates well with a recent observation that the links in a network are usually community--specific, while the nodes  might participate in different communities~\cite{ahn}.   

\acknowledgments
This research was partially supported by the U.S. ARO MURI grant W911NF--06--1--0094.
 A.E.A. would like to acknowledge support by Volkswagenstiftung.


\begin{thebibliography}{99}

\bibitem{newman} M. E. J. Newman, PNAS {\bf 103}, 8577 (2006). 

\bibitem{fortunato}S. Fortunato, Physics Reports {\bf 486}, 75--174 (2010).

\bibitem{danon} L. Danon {\it et al.}, J. Stat. Mech.: Theory Exp., P09008 (2005). 




\bibitem{leone} J. Reichardt and M. Leone, Phys. Rev. Lett. {\bf 101}, 078701 (2008).




\bibitem{zhu} X. Zhu and A.~B. Goldberg, {\em Introduction to Semi-Supervised Learning}, Synthesis Lectures
on Artificial Intelligence and Machine Learning, Morgan \& Claypool Publishers, (2009).

\bibitem{getz}G. Getz, N. Shental, and E. Domany, {\it Semi-supervised learning Ð a statistical physics approach} 
(Proc. 22nd ICML Workshop on Learning with Partially Classified Training Data, Bonn, Germany, 2005).

\bibitem{leone_sumedha} M. Leone {\it et al.}, Eur. Phys. J. B {\bf 66}, 125 (2008).


\bibitem{newman_weighted} M.~E.~J. Newman, Phys. Rev. E {\bf 70}, 056131 (2004).


\bibitem{karp} A. Condon and   R. M. Karp {\em Algorithms for Graph Partitioning on the Planted Partition Model}, Random Structures and Algorithms {\bf 18}, pp.116--140, 2001.

\bibitem{reich} J. Reichardt and S. Bornholdt, Phys. Rev. E {\bf 74} 016110 (2006). 


\bibitem{mezard} M. Mezard and G. Parisi, Europhys. Lett. {\bf 3}, 1067 (1987).

\bibitem{kanter}
I. Kanter and H. Sompolinsky, Phys. Rev. Lett. {\bf 58}, 164 (1987).

\bibitem{gold_rsb}Y. Y. Goldschmidt, Phys. Rev. B {\bf 43}, 8148 (1991).

\bibitem{mezard_rsb}M. Mezard and G. Parisi, Eur. Phys. J. B {\bf 20}, 217 (2001).

\bibitem{vb}L. Viana and A. J. Bray, J. Phys C, {\bf 18}, 3037 (1985).





\bibitem{ahn} Yong-Yeol Ahn, J. P. Bagrow, Sune Lehmann,  arXiv:0903.3178 (2009).



\bibitem{dvin}
Let us mention that the straightforward application (\ref{hatkhi}) of
the cavity method is related to assuming a single thermodynamic state
for the Gibbs distribution generated by the Hamiltonian (\ref{1.1})
\cite{vb,gold_rsb,mezard_rsb}. Though this assumption (also called a
replica symmetry assumption) produces reasonable estimates for extensive
quantities (such as the energy), it is normally violated whenever a
strong effect of frustrations is presented. However, even for strongly
frustrated spin-glass models the assumption on a single thermodynamic
state correctly predicts thresholds of second-order phase transitions \cite{vb}.
Because such thresholds are of the main interest in the present work,
and because going beyond of the assumption is notoriously
difficult (see, e.g., \cite{gold_rsb,mezard_rsb}), we restrict ourselves
to the above straightforward implementation of the cavity method. 



\end{thebibliography}
\end{document}